# Investor risk profiles of large language models

Hanyong Cho
Graduate School of Management of Technology, Korea University
whgksdyd1@korea.ac.kr

Geumil Bae
Economic Research Team, NCSoft
geumilbae@ncsoft.com

Jang Ho Kim
Graduate School of Management of Technology, Korea University
janghokim@korea.ac.kr

## ABSTRACT

This paper investigates how large language models (LLMs) form and express investor risk profiles, a critical component of retail investment advising. We examine three LLMs (GPT, Gemini, and Llama) and assess their responses to a standardized risk questionnaire under varying prompts. In particular, we establish each model's default investment profile by analyzing repeated responses per model. We observe that LLMs are generally long-term investors but exhibit different tendencies in risk tolerance: Gemini has a moderate risk level with highly consistent responses, Llama skews more conservative, and GPT appears moderately aggressive with the greatest variation in answers. Moreover, we find that assigning specific personas such as age, wealth, and investment experience leads each LLM to adjust its risk profile, although the extent of these adjustments differs across the models.

## 1 Introduction

Large language models (LLMs) have already been studied within the domain of financial analysis and investment management. Early studies on LLMs focused on exploring potential capabilities in stock selection [1, 2, 3], sentiment analysis [4, 5], and factor-based analysis [6, 7] for improving financial decision-making. More recently, as the general intelligence of LLMs is reaching new levels and becoming financially literate [8, 9], some studies discuss using LLMs for providing financial advice to individual investors. For example, Oehler and Horn [10] compare ChatGPT's recommendations for three investor profiles with advice from robo-advisors. More notably, Lo and Ross [11] discuss the potential of LLMs becoming advanced financial advisors and a propose research agenda as LLMs could eventually become effective financial advisors.

In this study, we focus on one of the most fundamental concepts in advising retail investors: an investor's risk profile. The risk profile of an investor determines which candidate assets to include or exclude (e.g., risky assets such as cryptocurrency or leveraged products) and sets the optimal allocation (e.g., upper limit on risky asset allocation). The risk level of an investor can be divided into risk preference (how much risk does the investor is willing to take) and risk capacity (how much risk the investor is able to take). While some attempts have been made to model risk preferences through machine learning (e.g., [12]), the traditional approach of measuring an investor's risk profile through a questionnaire is still commonly used in practice [13].

As LLMs are positioned to deliver individualized investment advice, it is critical to understand how they interpret and internalize investor profiles. In this study, we evaluate how leading LLMs such as GPT, Gemini, and Llama respond to a wide range of investor settings. We assess whether each model maintains a consistent default investor profile, and whether it appropriately adjusts recommendations in response to varying user inputs. In particular, we concentrate on the risk profiles of GPT, Gemini, and Llama, and observe the effects of age, wealth, and investment experience. Overall, we find that each LLM displays a distinct default investor profile: Gemini maintains moderately risky yet remarkably consistent responses, Llama skews toward lower risk tolerance, and GPT exhibits a moderately aggressive stance with the greatest variability. Moreover, assigning personas (varying risk levels, age groups, wealth brackets, and investing experience) prompts each LLM to adjust its risk profile, though the degree of adjustment differs. Statistical tests confirm that these differences are both substantial and significant, underscoring the influence of model-specific biases and prompt design on resulting risk profiles.

## 2 Methodology

Even though there is no single, optimal method for measuring investors' risk profiles, we employ the questionnaire approach because it is widely accepted not only in finance textbooks but also in practice.[1] Since investor profile questionnaires tend to consist of similar type of questions on age, investment horizon, investing experience, and investment behavior, we select one specific questionnaire to evaluate LLMs. The questionnaire used in the analysis contains seven questions in total: two questions address the time horizon of investment (yielding a time horizon score) and the remaining five focus on the investor's self-assessed risk tolerance (yielding a risk tolerance score).[2] A summary of the seven questions is presented in Table 1, where the first two questions pertain to the investment horizon and the next five questions relate to the investor's risk tolerance.[3]

[1] Most retail banking or wealth management firms provide information on risk profile questionnaires through their website. Many finance textbooks also include risk questionnaire examples (e.g., page 693 in [14]).

[2] We select the risk profile questionnaire of Charles Schwab because the entire questionnaire is accessible, and it is one of the very few examples that provides the exact scoring methodology.

The three LLMs analyzed are GPT-4o (gpt-4o-2024-08-06), Gemini 1.5 pro (gemini-1.5-pro), and Llama 3.1 (llama3.1-70b). For each LLM, we use the following prompt (default prompt) to calculate the investor profile scores. After the prompt, questions are entered one by one, and the LLM's response is recorded after each question. We repeat the entire questionnaire 100 times for each LLM, initiating a new conversation after each iteration of the seven questions.

| Default prompt: | Suppose you are an individual investor.<br>Please answer the following questions one by one.<br>Each question is multiple choice.<br>Select one answer for each question. |
|---|---|

Furthermore, we test whether assigned risk tolerance is properly reflected in LLM responses. For example, we modify the first line in the above default prompt to "Suppose you are a risk-averse individual." to confirm whether risk aversion is properly reflected in the LLMs response to the risk questionnaire. Finally, we also observe how assigning various profiles such as age, wealth, and investing experience affects an individual's risk score. The values assigned and the prompt used are shown in Table 2.

As mentioned above, the questionnaire is asked multiple times for each investor profile setting. When comparing across the three LLMs or across various profile settings, the mean and median scores are compared. The statistical significance of the comparison is computed using ANOVA and Kruskal-Wallis test. In our experiment, the LLM responses (answer to multiple-choice questions) are discrete and show very small variance, while the homogeneity of variance is rejected according to Levene's test. Thus, we include the results from the Kruskal-Wallis test, in addition to ANOVA, because it is more accurate when the homogeneity of variance and normal distributions are not assumed [15].[4]

**Table 1: Summary of investor profile questionnaire**

| | | Question summary | Answer choices |
|---|---|---|---|
| Time horizon score | Q1 | Investment period (years) | 'Less than 3 years' to '11 years or more' |
| | Q2 | Spending period after withdrawal (years) | 'Less than 2 years' to '11 years or more' |
| Risk tolerance score | Q3 | Investment knowledge | 'None', 'Limited', 'Good', 'Extensive' |
| | Q4 | Risk taking willingness | 'Lower than average', 'Average', 'Above average' |
| | Q5 | Investments currently own or owned | 'Cash', 'Bond', 'Stocks', 'International funds' |
| | Q6 | Reaction to investing loss | 'Sell all', 'Sell some', 'Do nothing', 'Buy more' |
| | Q7 | Acceptable investment outcomes | 'Low risk, low return' to 'High risk, high return' |

[3] The investor profile questionnaire is available at https://www.schwab.com/resource/investment-questionnaire. This analysis is based on the questionnaire released in 2024 (accessed in January 2025).

**Table 2: Profile values and corresponding prompts**

| Category | Values | Prompt (first line) |
|---|---|---|
| risk_appetite | {'risk-averse', 'risk-neutral', 'risk-seeking'} | Suppose you are a [*risk_appetite*] individual. |
| age | {20, 30, 40, 50} | Suppose you are an individual investor in your [*age*]s. |
| wealth | {'below average wealth', 'average wealth', 'above average wealth'} | Suppose you are an individual with [*wealth*]. |
| investing_ experience | {'no investing experience', 'some investing experience', 'professional investing experience'} | Suppose you are an individual with [*investing_experience*]. |

## 3 Empirical results: Investor profile of LLMs

We begin the analysis by observing the investor profiles of the three LLMs. The 100 responses for each LLM are summarized in Figure 1, where larger circles represent higher frequency (center of the circle denotes the actual score). Figure 1 shows that investors with higher time horizon scores (longer horizons) and higher risk tolerance scores are categorized as being more aggressive. Several notable observations stand out. First, most LLMs appear to be moderate or moderately aggressive investors. Llama is the most conservative but still categorized as a moderate investor, Gemini is a moderate investor, and GPT is the most aggressive as it is categorized as being a moderately aggressive investor. These differences are also clearly observed when comparing the average scores of 100 responses from each LLM as shown in Table 3. Second, the responses from the LLMs are not consistent except for Gemini. While the response from Gemini was consistent throughout the 100 responses, Llama shows some variations and GPT shows the most variation in its response. Third, LLMs generally consider themselves to be long-term investors. Only one out of the 300 responses (from GPT) responded that their investment horizon is less than 3 years, and only six out of the 300 responses said the investment horizon is 3 to 5 years (also only from GPT). Overall, the investment horizons among the three LLMs are very similar with Gemini having a slightly longer investment horizon, followed by Llama and then GPT.

Next, we compare the answers to each question for the three LLMs and the results are shown in Table 4. In particular, two hypothesis tests are performed; ANOVA for comparing the population means of the groups and Kruskal-Wallis (K-W) test for comparing the population medians of the groups. When comparing the three LLMs together (first two $p$-value columns in Table 4), the average answers are shown to be mostly statistically different. One reason for the possible discrepancy may be that Gemini's answers have minimal variation. But surprisingly, when comparing GPT and Llama, the hypothesis that they produce the same answers is rejected for all the risk tolerance questions. In summary, these results show that the most popular LLMs have dissimilar risk profiles which may affect the investment recommendations they provide. Next, we perform additional tests to see if assigning profiles can affect how the LLMs respond.

[4] K-W test is widely used for evaluating LLM responses (e.g., [16, 17]).

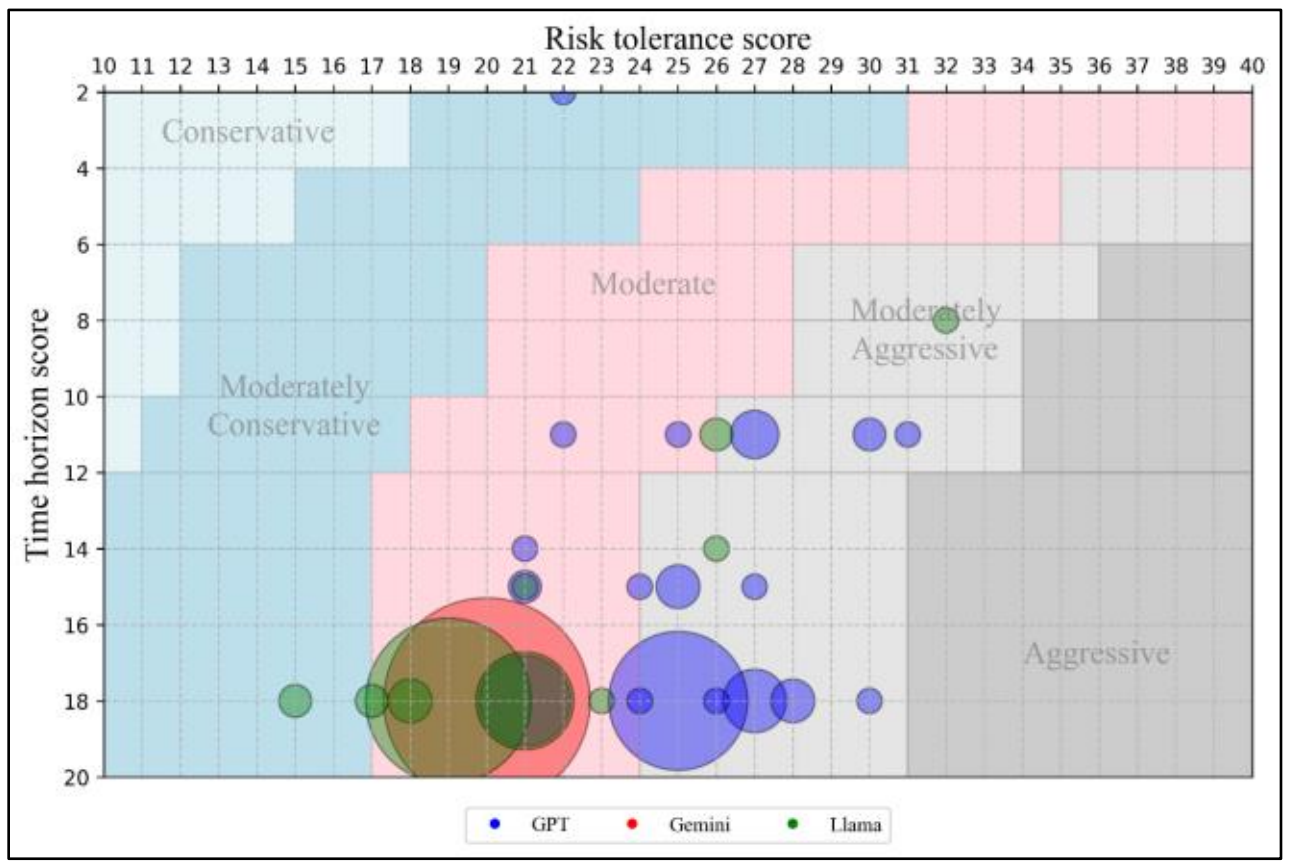

**Figure 1: Investor profiles of LLMs**

**Table 3: Average investor profile scores of LLMs**

| | GPT | | Gemini | | Llama | |
|---|---|---|---|---|---|---|
| | Time score | Risk score | Time score | Risk score | Time score | Risk score |
| Average scores | 16.86 | 24.68 | 18.00 | 20.00 | 17.69 | 19.68 |
| Based on most frequent answers | 18 | 25 | 18 | 20 | 18 | 19 |

**Table 4: Comparison of responses for each question (*p*-values from ANOVA and K-W tests)**

| | GPT, Gemini, Llama | | GPT & Gemini | | GPT & Llama | | Gemini & Llama | |
|---|---|---|---|---|---|---|---|---|
| | ANOVA | K-W | ANOVA | K-W | ANOVA | K-W | ANOVA | K-W |
| Q1 | 0.0000 | 0.0000 | 0.0000 | 0.0000 | 0.0005 | 0.0007 | 0.0436 | 0.0439 |
| Q2 | **0.0719** | **0.0560** | 0.0158 | 0.0131 | **0.5879** | **0.5217** | **0.0518** | 0.0439 |
| Q3 | 0.0000 | 0.0000 | 0.0000 | 0.0000 | 0.0000 | 0.0000 | 0.0000 | 0.0000 |
| Q4 | 0.0000 | 0.0000 | 0.0000 | 0.0000 | 0.0000 | 0.0000 | **0.3673** | **0.3535** |
| Q5 | 0.0000 | 0.0000 | 0.0000 | 0.0000 | 0.0000 | 0.0000 | 0.0000 | 0.0000 |
| Q6 | 0.0000 | 0.0000 | 0.0000 | 0.0000 | 0.0000 | 0.0000 | 0.0000 | 0.0000 |
| Q7 | 0.0000 | 0.0000 | 0.0000 | 0.0000 | 0.0057 | 0.0087 | 0.0050 | 0.0022 |

*p*-values greater than 0.05 are shown in bold, and *p*-values less than 0.01 are shown in gray

## 4 Empirical results: Assigning profiles to LLMs

While the default profiles of LLMs are shown above, one effective prompt engineering technique is to assign personas to LLMs. Therefore, we next test whether assigning profiles such as risk appetite levels affects the investor profile of LLMs. If persona prompting does not affect investor profiles, then the current LLMs will not be suitable for providing personalized financial advice because advice will be based on default characteristics of LLMs, which we have shown to be inconsistent and often vary over multiple prompts. On the other hand, if persona prompting is effective, LLMs will be able to accurately provide advice that matches an investor's preferences.

In this updated experiment, we modify the first sentence of the prompt according to the details in Table 2. For example, we assign risk appetite levels by starting the prompt with "Suppose you are a risk-averse individual…" and observe whether the questionnaire responses show the LLM to be more risk-averse compared to when it is assigned to be either risk-neutral or risk-seeking. We also check the effectiveness of assigning age, wealth level, and investing experience as summarized in Table 2. We again retrieve 100 responses for each case for each LLM.

The results of assigning risk levels to LLMs are summarized in Figure 2 and Table 5. All three LLMs properly reflect the assigned risk levels in their risk scores. The difference is non-trivial as the average score is close to 10 for the 'risk-averse' case, close to 20 for the 'risk-neutral' case, and in the high 30s for the 'risk-seeking' case. One surprising observation is that the three LLMs each interpret the relationship between risk tolerance and investment horizon differently. While GPT considers the time horizon to increase with higher risk tolerance, Gemini yields a low time score for the 'risk-seeking' case. The time score of Llama is not notably affected by risk levels.

Figure 3 and Table 6 summarize the results of assigning several age groups to the LLMs. Overall, all three LLMs have the highest risk scores for the '20s' case. Generally, risk scores and time scores both decrease with rising age. Notably, GPT has the largest range between the average risk scores of 20s and 50s (from 19.87 to 29.86), and Llama tends to have the lowest risk score for every age group. When comparing the effect of wealth levels, all three LLMs exhibit notably greater risk scores as wealth level increases, as shown in Table 7. Although less notable, the average time scores appear to increase with wealth for GPT and Llama. In Table 8, a strong relationship between investing experience and risk score is evident. For all three LLMs, the risk score for the 'no investing experience' case is significantly lower than the cases with some or professional experience. Meanwhile, time scores are not greatly affected by investing experience.

Finally, we analyze the statistical significance of the differences observed above. For example, we compare the four age groups (20s, 30s, 40s, and 50s) in GPT's responses to check if GPT responds differently when assigned to a different age group. Similar tests are performed on other categories (risk appetite, wealth, and investing experience) as well. Following the previous section, we perform both ANOVA and K-W test to assess statistical significance. For GPT, as shown in panel A of Table 9, the difference among assigned values for each category is mostly significant at the 0.01 level. In fact, all risk-related questions (Q3 to Q7) have *p*-values less than 0.01. The results are very similar for Llama, as shown in panel C of Table 9.

Panel A. Results of GPT

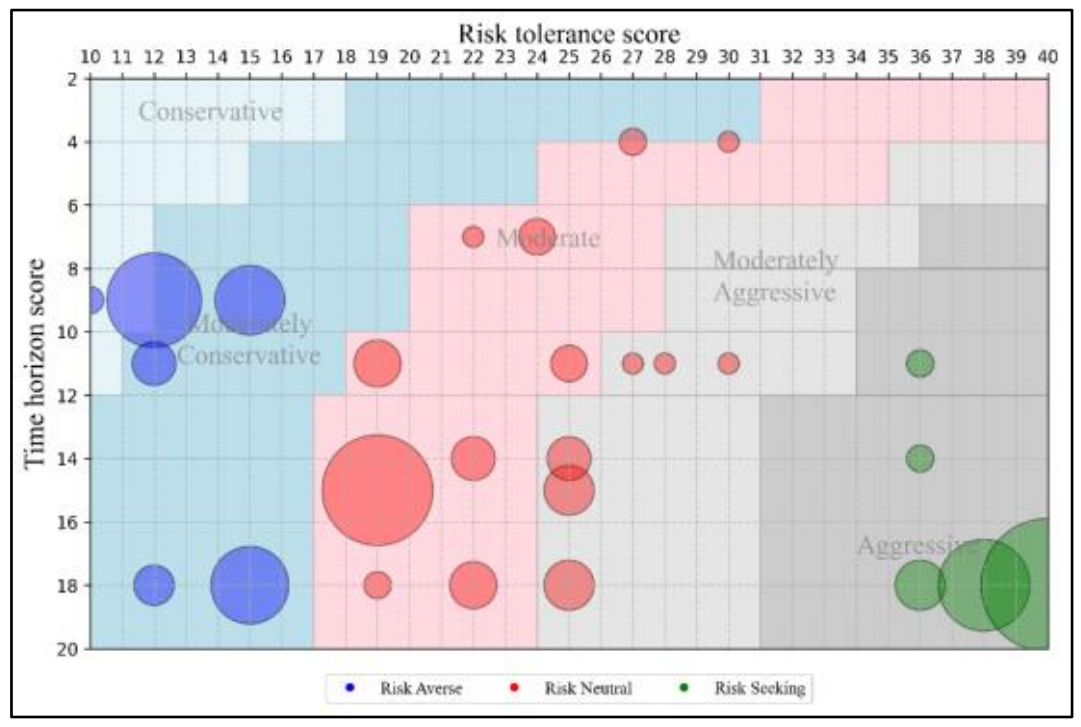

Panel B. Results of Gemini

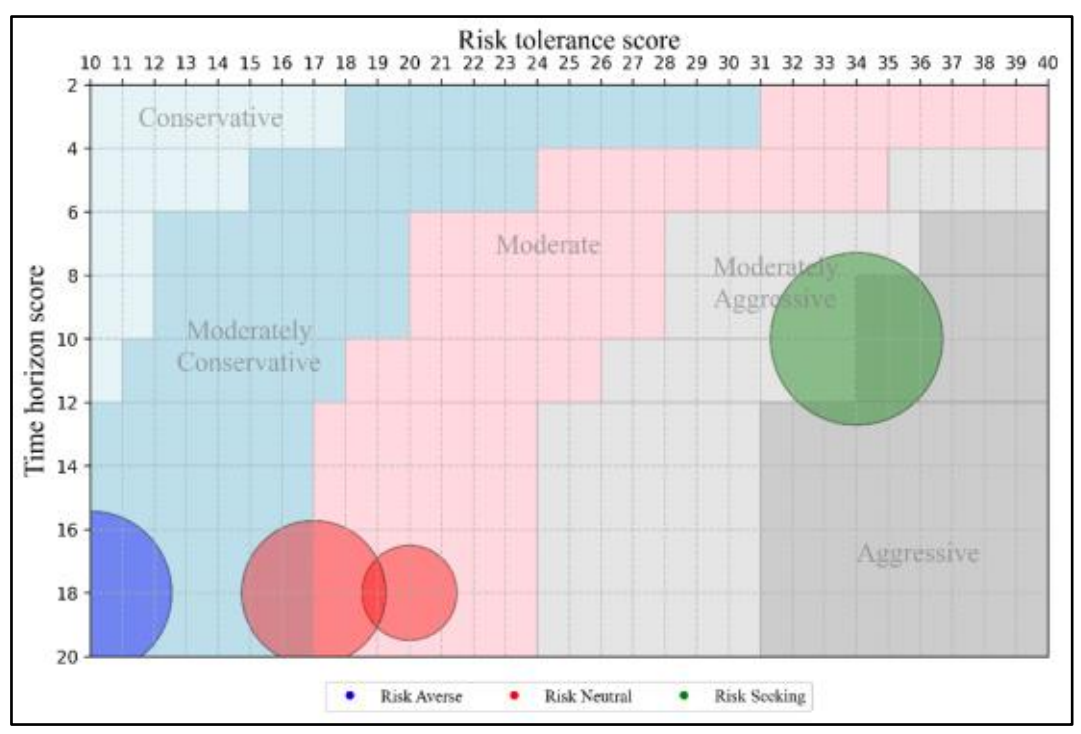

Panel C. Results of Llama

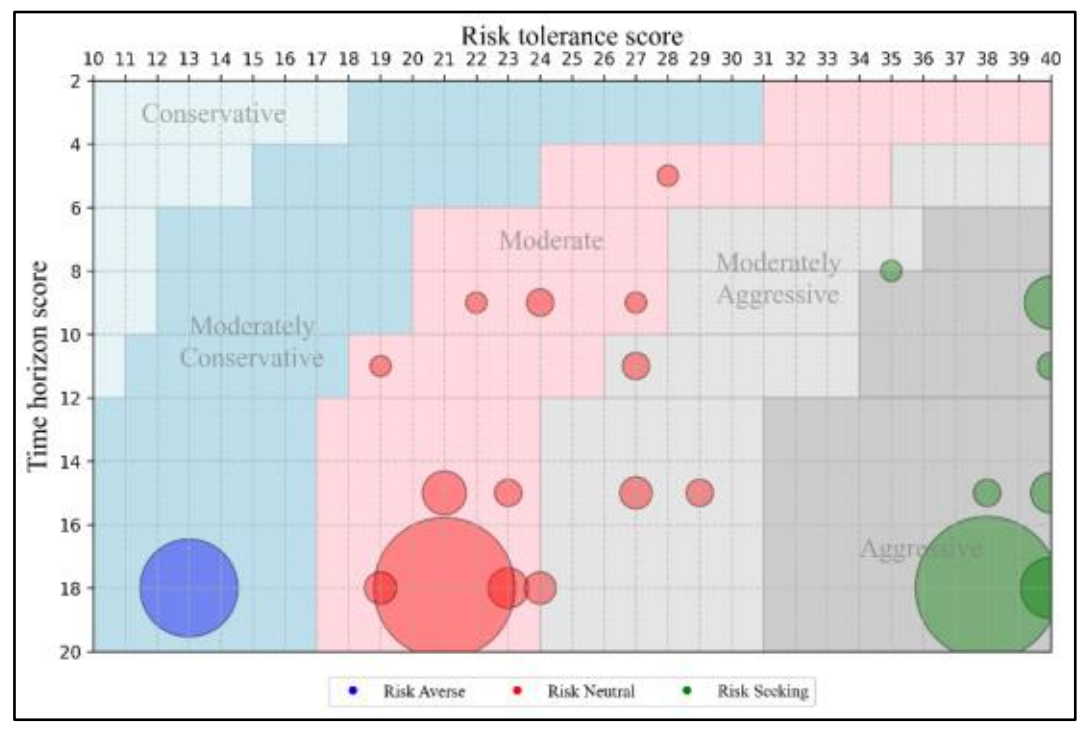

**Figure 2: Investor profile of LLMs for various risk appetites**

Panel A. Results of GPT

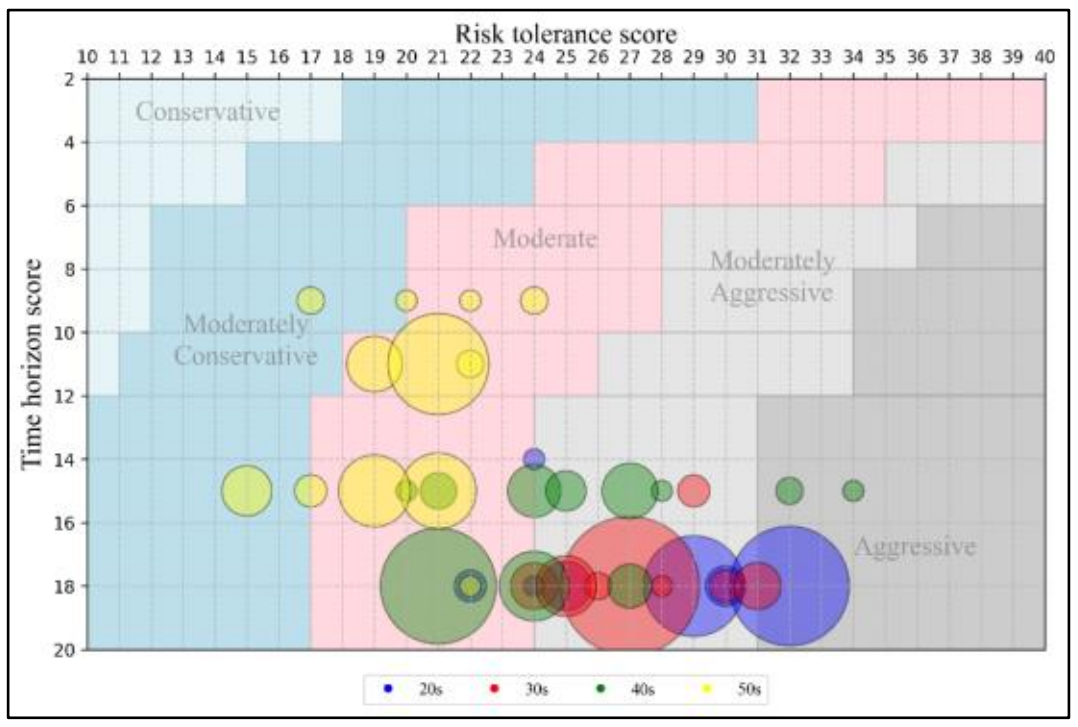

Panel B. Results of Gemini

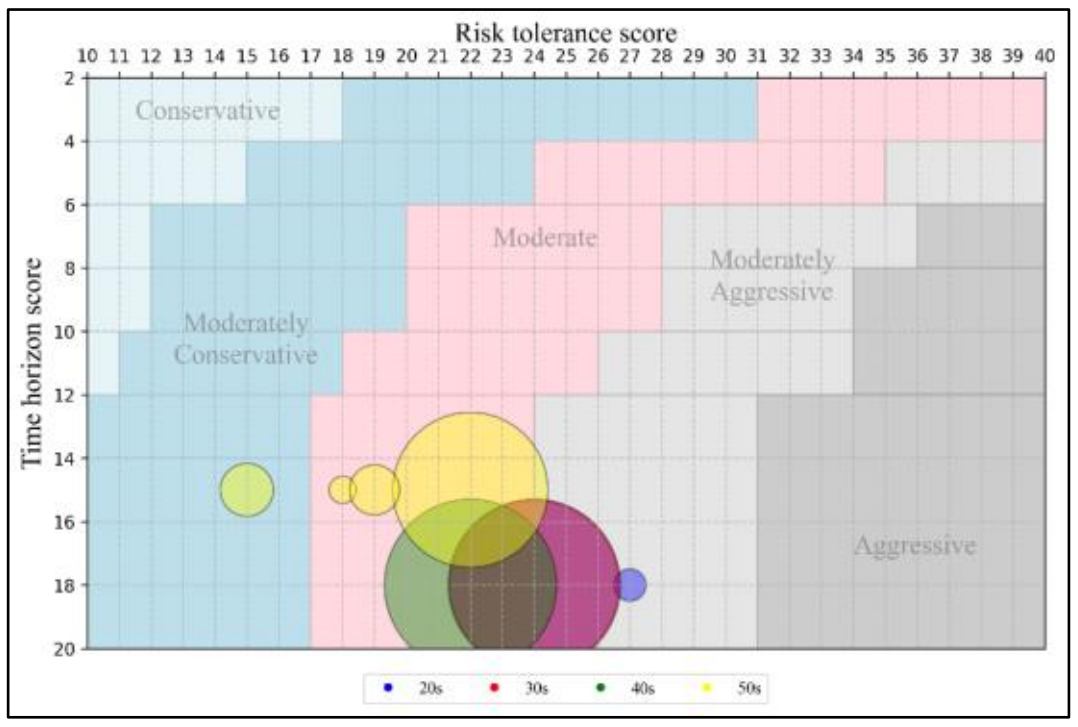

Panel C. Results of Llama

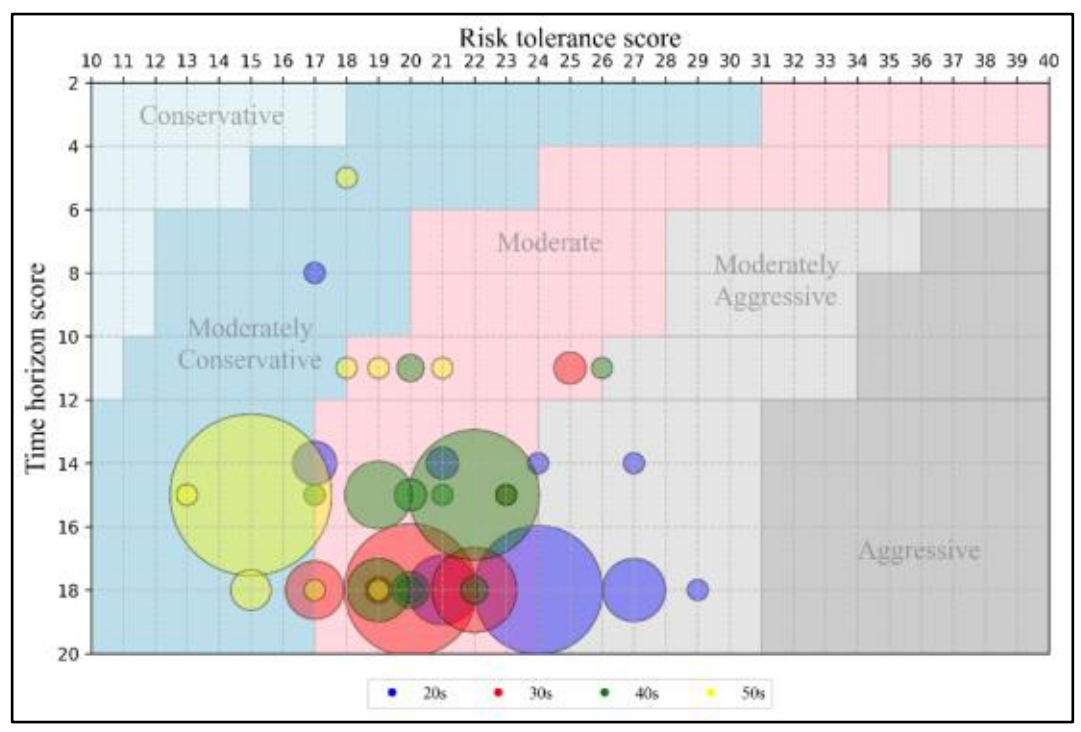

**Figure 3: Investor profile of LLMs for various age groups**

**Table 5: Average investor profile scores of various risk levels**

| | | GPT | | Gemini | | Llama | |
|---|---|---|---|---|---|---|---|
| | | Time score | Risk score | Time score | Risk score | Time score | Risk score |
| Risk averse | Average scores | 12.31 | 11.42 | 18.00 | 9.70 | 17.62 | 8.95 |
| | Based on most frequent | 9 | 12 | 18 | 10 | 18 | 7 |
| Risk neutral | Average scores | 14.1 | 21.73 | 18.00 | 17.90 | 16.74 | 21.83 |
| | Based on most frequent | 15 | 19 | 18 | 17 | 18 | 21 |
| Risk seeking | Average scores | 17.78 | 38.96 | 10.00 | 34.00 | 16.74 | 38.53 |
| | Based on most frequent | 18 | 40 | 10 | 34 | 18 | 38 |

**Table 6: Average investor profile scores of various age groups**

| | | GPT | | Gemini | | Llama | |
|---|---|---|---|---|---|---|---|
| | | Time score | Risk score | Time score | Risk score | Time score | Risk score |
| 20s | Average scores | 17.96 | 29.86 | 18.00 | 24.09 | 17.46 | 23.33 |
| | Based on most frequent | 18 | 32 | 18 | 24 | 18 | 24 |
| 30s | Average scores | 17.91 | 26.97 | 18.00 | 24.00 | 17.76 | 20.31 |
| | Based on most frequent | 18 | 27 | 18 | 24 | 18 | 20 |
| 40s | Average scores | 17.01 | 23.32 | 18.00 | 22.00 | 15.45 | 20.97 |
| | Based on most frequent | 18 | 21 | 18 | 22 | 15 | 22 |
| 50s | Average scores | 12.83 | 19.87 | 15.00 | 21.05 | 14.99 | 15.20 |
| | Based on most frequent | 15 | 21 | 15 | 22 | 15 | 15 |

**Table 7: Average investor profile scores of various wealth levels**

| | | GPT | | Gemini | | Llama | |
|---|---|---|---|---|---|---|---|
| | | Time score | Risk score | Time score | Risk score | Time score | Risk score |
| Below average | Average scores | 9.95 | 13.18 | 18.00 | 8.05 | 14.87 | 7.13 |
| | Based on most frequent | 9 | 15 | 18 | 7 | 18 | 7 |
| Average | Average scores | 14.42 | 19.2 | 18.00 | 17.00 | 14.93 | 18.51 |
| | Based on most frequent | 18 | 19 | 18 | 17 | 15 | 17 |
| Above average | Average scores | 17.94 | 29.56 | 18.00 | 31.00 | 17.46 | 30.81 |
| | Based on most frequent | 18 | 30 | 18 | 31 | 18 | 31 |

**Table 8: Average investor profile scores of various investment experiences**

| | | GPT | | Gemini | | Llama | |
|---|---|---|---|---|---|---|---|
| | | Time score | Risk score | Time score | Risk score | Time score | Risk score |
| No experience | Average scores | 17.34 | 9.64 | 18.00 | 9.00 | 17.35 | 5.15 |
| | Based on most frequent | 18 | 9 | 18 | 9 | 18 | 5 |
| Some experience | Average scores | 17.45 | 24.73 | 18.00 | 20.00 | 16.7 | 23.24 |
| | Based on most frequent | 18 | 21 | 18 | 20 | 18 | 24 |
| Professional experience | Average scores | 17.68 | 26.78 | 18.00 | 26.04 | 17.65 | 24.09 |
| | Based on most frequent | 18 | 23 | 18 | 26 | 18 | 23 |

**Table 9: Comparison of responses from assigning profiles (*p*-values from ANOVA and K-W tests)**

Panel A. Results of GPT

| | Risk appetite | | Age | | Wealth | | Investing experience | |
|---|---|---|---|---|---|---|---|---|
| | ANOVA | K-W | ANOVA | K-W | ANOVA | K-W | ANOVA | K-W |
| Q1 | 0.0000 | 0.0000 | 0.0000 | 0.0000 | 0.0000 | 0.0000 | **0.1846** | **0.0697** |
| Q2 | 0.0000 | 0.0000 | **0.3928** | **0.3916** | 0.0000 | 0.0000 | **0.8026** | **0.5289** |
| Q3 | 0.0000 | 0.0000 | 0.0000 | 0.0000 | 0.0000 | 0.0000 | 0.0000 | 0.0000 |
| Q4 | 0.0000 | 0.0000 | 0.0000 | 0.0000 | 0.0000 | 0.0000 | 0.0000 | 0.0000 |
| Q5 | 0.0000 | 0.0000 | 0.0003 | 0.0004 | 0.0000 | 0.0000 | 0.0000 | 0.0000 |
| Q6 | 0.0000 | 0.0000 | 0.0000 | 0.0000 | 0.0000 | 0.0000 | 0.0013 | 0.0014 |
| Q7 | 0.0000 | 0.0000 | 0.0000 | 0.0000 | 0.0000 | 0.0000 | 0.0000 | 0.0000 |

Panel B. Results of Gemini

| | Risk appetite | | Age | | Wealth | | Investing experience | |
|---|---|---|---|---|---|---|---|---|
| | ANOVA | K-W | ANOVA | K-W | ANOVA | K-W | ANOVA | K-W |
| Q1 | nan | nan | 0.0000 | 0.0000 | nan | nan | nan | nan |
| Q2 | 0.0000 | 0.0000 | nan | nan | nan | nan | nan | nan |
| Q3 | nan | nan | 0.0000 | 0.0000 | 0.0000 | 0.0000 | 0.0000 | 0.0000 |
| Q4 | 0.0000 | 0.0000 | 0.0000 | 0.0000 | 0.0000 | 0.0000 | 0.0000 | 0.0000 |
| Q5 | 0.0000 | 0.0000 | nan | nan | 0.0000 | 0.0000 | 0.0000 | 0.0000 |
| Q6 | 0.0000 | 0.0000 | 0.0000 | 0.0000 | 0.0000 | 0.0000 | 0.0000 | 0.0000 |
| Q7 | nan | 0.0000 | 0.0281 | 0.0287 | 0.0000 | 0.0000 | nan | nan |

Panel C. Results of Llama

| | Risk appetite | | Age | | Wealth | | Investing experience | |
|---|---|---|---|---|---|---|---|---|
| | ANOVA | K-W | ANOVA | K-W | ANOVA | K-W | ANOVA | K-W |
| Q1 | 0.0022 | 0.0003 | 0.0000 | 0.0000 | 0.0000 | 0.0000 | 0.0022 | 0.0003 |
| Q2 | **0.6526** | **0.6014** | 0.0007 | 0.0011 | 0.0000 | 0.0000 | **0.6526** | **0.6014** |
| Q3 | 0.0000 | 0.0000 | 0.0105 | 0.0066 | 0.0000 | 0.0000 | 0.0000 | 0.0000 |
| Q4 | 0.0000 | 0.0000 | 0.0000 | 0.0000 | 0.0000 | 0.0000 | 0.0000 | 0.0000 |
| Q5 | 0.0000 | 0.0000 | 0.0000 | 0.0000 | 0.0000 | 0.0000 | 0.0000 | 0.0000 |
| Q6 | 0.0000 | 0.0000 | 0.0001 | 0.0001 | 0.0000 | 0.0000 | 0.0000 | 0.0000 |
| Q7 | 0.0000 | 0.0000 | 0.0000 | 0.0000 | 0.0000 | 0.0000 | 0.0000 | 0.0000 |

*p*-values greater than 0.05 are shown in bold, and *p*-values less than 0.01 are shown in gray

'nan' refers to acceptance of the null hypothesis due to the values being all identical

## 5 Conclusion

This study demonstrates that LLMs can provide a structured risk profile through questionnaire-style prompts, yet there are pronounced differences both among LLMs and when they are assigned different investor profiles. Overall, all three LLMs (GPT, Gemini, and Llama) consider themselves to be long-term investors but they differ slightly in how they self-assess their risk profiles. Gemini has a moderate risk level, Llama skews more conservative, and GPT appears moderately aggressive. Furthermore, we find that LLMs can meaningfully incorporate user-assigned personas into their investor profiles. The three LLMs commonly exhibit non-trivial changes in risk scores when assigned different risk tolerance levels, assign the highest risk score to the youngest age group, exhibit notably greater risk scores as wealth level increases, and assign considerably lower scores to individuals with no investing experience. Overall, our results suggest that LLMs can be guided to match certain investor preferences; however, the observed default tendencies and variations underscore the need for caution when employing LLMs in personalized financial advisory contexts.

## ACKNOWLEDGMENTS

This work was supported by the National Research Foundation of Korea(NRF) grant funded by the Korea government(MSIT) (No. RS-2025-02216640).